\documentclass[a4paper,fleqn]{cas-dc}

\usepackage[numbers]{natbib}

\def\tsc#1{\csdef{#1}{\textsc{\lowercase{#1}}\xspace}}
\tsc{WGM}
\tsc{QE}

\begin{document}
\let\WriteBookmarks\relax
\def\floatpagepagefraction{1}
\def\textpagefraction{.001}

\shorttitle{A Cascade Superradiance Model}    

\shortauthors{G.O. Ariunbold}  


\title [mode = title]{A Cascade Superradiance Model} 



%

\author[1]{Gombojav O. Ariunbold}[orcid=0000-0003-0430-7256]

\cormark[1]

\fnmark[1]

\ead{ag2372@msstate.edu}

\ead[url]{www.ariunbold.physics.msstate.edu}


\affiliation[1]{organization={Department of Physics and Astronomy},
            addressline={Mississippi State University}, 
            city={Starkville},
            postcode={39762}, 
            state={MS},
            country={USA}}







\cortext[1]{Corresponding author}



\begin{abstract}
Intriguing collective spontaneous cascade emissions have recently been realized. In despite of much success, a depth understanding of the complexity is still lacking. With this motivation, a new simple cascade superradiance model is developed in this work. The existing model of identical two-level atoms is reexamined with a new insight. Temporal evolutions of average time delays and the fluctuations are introduced and the superradiance time delays are obtained in four different ways. These formulations allow to extend the two-level model to a cascade three-level model. The correlated two-mode emissions and the characteristics are discussed in detail. In the future, the correlated emissions from the collective atoms may be used, for example, in quantum noise quenching.
\end{abstract}


\begin{highlights}
\item Temporal evolutions of average time delays and the fluctuations are introduced.
\item The well-known superradiance time delays are obtained in four different ways.
\item A new cascade superradiance model for two-mode three-level atoms is introduced and studied.
\item The characteristics of the spontaneously formed correlated two-mode cooperative emissions are obtained.

\end{highlights}

\maketitle
%

\section{Introduction}\label{6}



%
The collective spontaneous emissions from the identical two-level atomic ensemble was first introduced by Robert Dicke in 1954~\cite{dicke}. Since then, the Dicke superradiance (SR) has brought a tremendous attention to the quantum optics community~\cite{rehler, BSH1, BSH2, degiorgio, lee, agar, feld1, feld2, marek, vrehen, vrehen2, gross, book} . However, the recent experimental realizations of the complex quantum systems of strongly correlated many bodies in gas, liquid and solid-state phases~\cite{ralf,ari1,ariOL,ariSok,ariFluct,bohnet,ariNa,ariBack,raino,diamond,calcium,xenon2,braggio,ariPyr,chiossi,findik,biliroglu,ariCs} demand the understanding of collective phenomena in much broader sense~\cite{marlan,alexey,kochar,gundogdu}. For example, the surprisingly tolerant superfluorescence (SF) has been recently demonstrated in perovskite materials at room temperature~\cite{findik,biliroglu}, which may be understood by a quantum analog of vibration isolation mechanism~\cite{gundogdu}.

This work is inspired by the two recent experiments~\cite{chiossi, braggio}. In 2020, Braggio et al.~\cite{braggio} succeeded to generate the macroscopic coherent states consisting of trillions of atoms in a cryogenically cooled rare-earth doped material by an incoherent pumping, to our surprise, using the continuous wave (CW) laser. In 2021, using the same incoherent pumping technique in the same material, Chiossi et al.~\cite{chiossi} observed the cascade superfluorescence (CSF). Incoherently pumped CSF is particularly remarkable to produce: i) the cleanest CSF and ii) a system with a large number of macroscopically coupled atoms for extended time. In the future, these techniques can be applied, e.g., to reach the mHz linewidth laser~\cite{meisner,meisner2,strontium2}. On contrary, the early works on the CSF~\cite{gross2,okada, okada2, ikeda2,strontium,florian,xenon} and yoked SF~\cite{brownell, ari1,ariOL,ariFluct,ariNa,ariBack} actively involved an on-resonant  coherent pumping, e.g., using the ultrashort laser pulses. In this work, a cascade superradiance (CSR) model is presented as an extension of the six-decade old simple two-level model~\cite{rehler, BSH1, BSH2, lee, agar}. These models are rather simple in the sense that no extended medium is considered and solely atomic operators are used in the master equations derived in~\cite{lee, agar}. The present model is referred to as CSR rather than CSF because in our case the atoms are excited coherently. Although, the CSR involves coherent pumping preparation from the off-resonant ground state, but that ground state is not part of the SR emissions. This arrangement also removes the early CSF and yoked SF considerations from the model. Therefore, the CSR is a straightforward extension of Dicke SR. Nevertheless, the CSR and recently demonstrated CSF have some common characteristics. The most intriguing property of CSF is that a pair SF pulses is generated at random phase however the relative phase between the pulses are correlated~\cite{chiossi}. The SF time delay and the other SF characteristics for the second pulse can be achieved in spite of the CW laser pumping (with no phase information). Similarly, in this work, the pair pulses generation and CSR characteristics are obtained.

\section{A Simple Superradiance Model: Revisited}
We begin with a system of identical two-level $N$ atoms confined to a small volume compared to the wavelength of the emitted radiations. As derived in~\cite{book, BSH1, BSH2, lee, agar}, the equation of motion can be rewritten as
%
\begin{eqnarray} \label{main2level}
\frac{dP_{n}(t)}{d t} =  I({n+1})P_{n+1}(t)-I({n})P_{n}(t) 
\end{eqnarray}
where $P_{n}(t)$ is probability for a system with $n$ atoms in the upper level and $N-n$ atoms in the lower level for an atomic state $|n,N-n\rangle$ at time $t$. 
The so-called cooperative decay rate~\cite{lee,agar} are given by $I({n})=n(N-n+1)$ and $\Gamma$ is the spontaneous decay rate for the excited state of a single two-level atom. $\mu$ is a geometrical factors for a pencil shape sample volume~\cite{feld2} and a dimensionless time $t$ is introduced as $t \rightarrow \mu\Gamma t$. The sum of the probability $P_{n}(t)$ is independent on
time and must be always one $\sum_{n=0}^{N} P_{n}(t) \equiv 1$.
\begin{figure} [!ht]   
\begin{center}
\includegraphics[width=60mm]{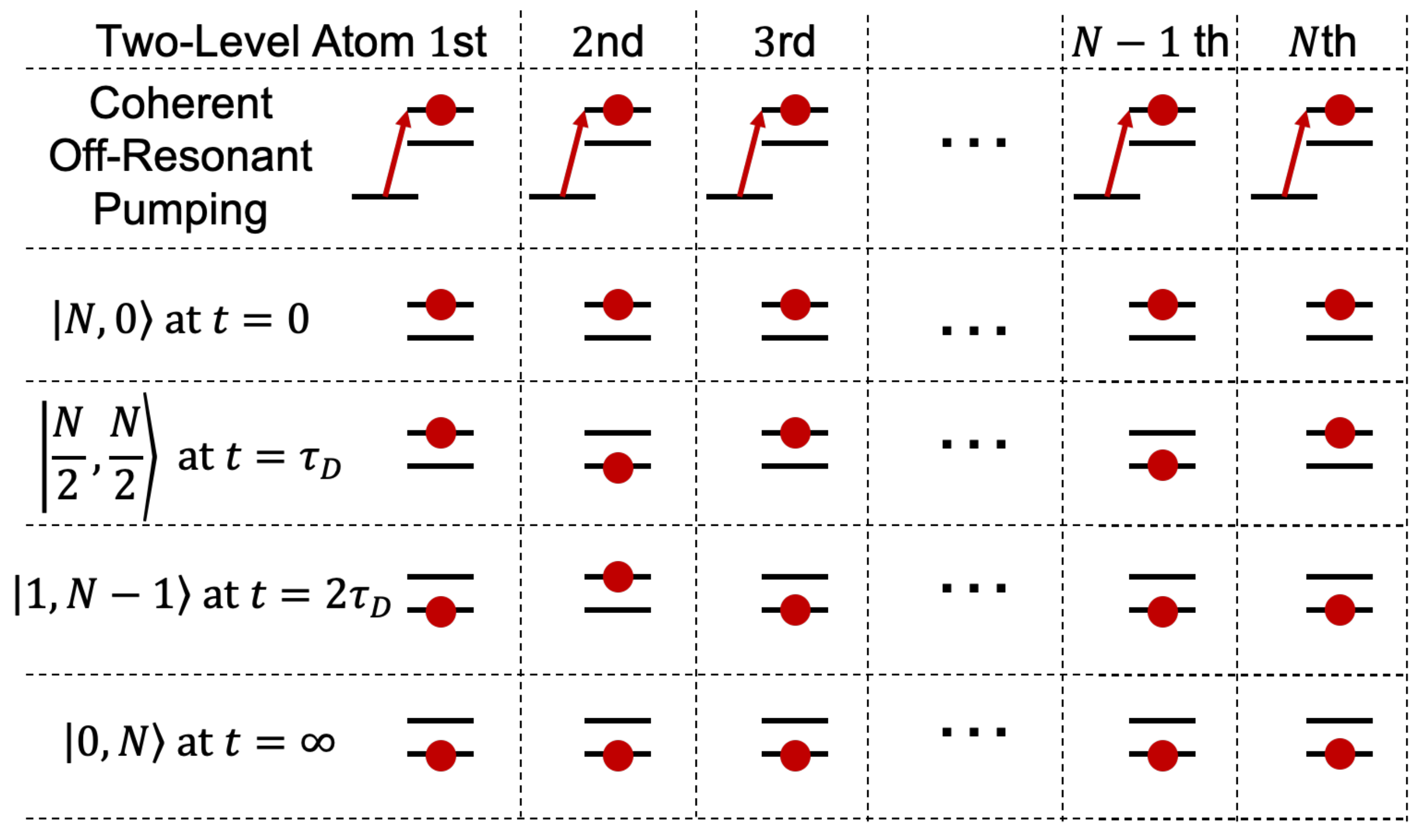}
\caption{Collective spontaneous emissions: A pictorial explanation of Dicke superradiance.} 
\label{sketch}
\end{center}
\end{figure}
%
%
This model is sketched in Fig.~\ref{sketch}. The coherent but off-resonant laser pumping prepares $N$ two-level atoms in the upper level from the (off-resonant) ground state at a time scale much shorter than duration of any atomic decay processes (e.g., spontaneous emissions, dephasing) to the lower level (second row). This is how all atoms are initially prepared in the upper level at $t=0$ and, hence, none is in the lower level, $P_{n}(0)=\delta_{n,N}$ (third row). The SR emissions reach the maximum at $t=\tau_D$ (SR time delay), where a half of atoms is in the upper level and other half is in the lower level (fourth row). As is shown later, the atoms are prepared in a special state at $t=2\tau_D$, where statistically a single atom is excited in the upper level while the rest is in the lower level. Therefore, the atomic system as a whole is  prepared as a precondition for a single-photon Dicke SR (fifth row). Later ($t=\infty$) all atoms fall down in the lower level while none is left anymore in the upper level $P_{n}(\infty)=\delta_{n,0}$ (sixth row).
%
%
%

%
%
\begin{figure}[!ht] \label{2level}
\begin{center}
\includegraphics[width=80mm]{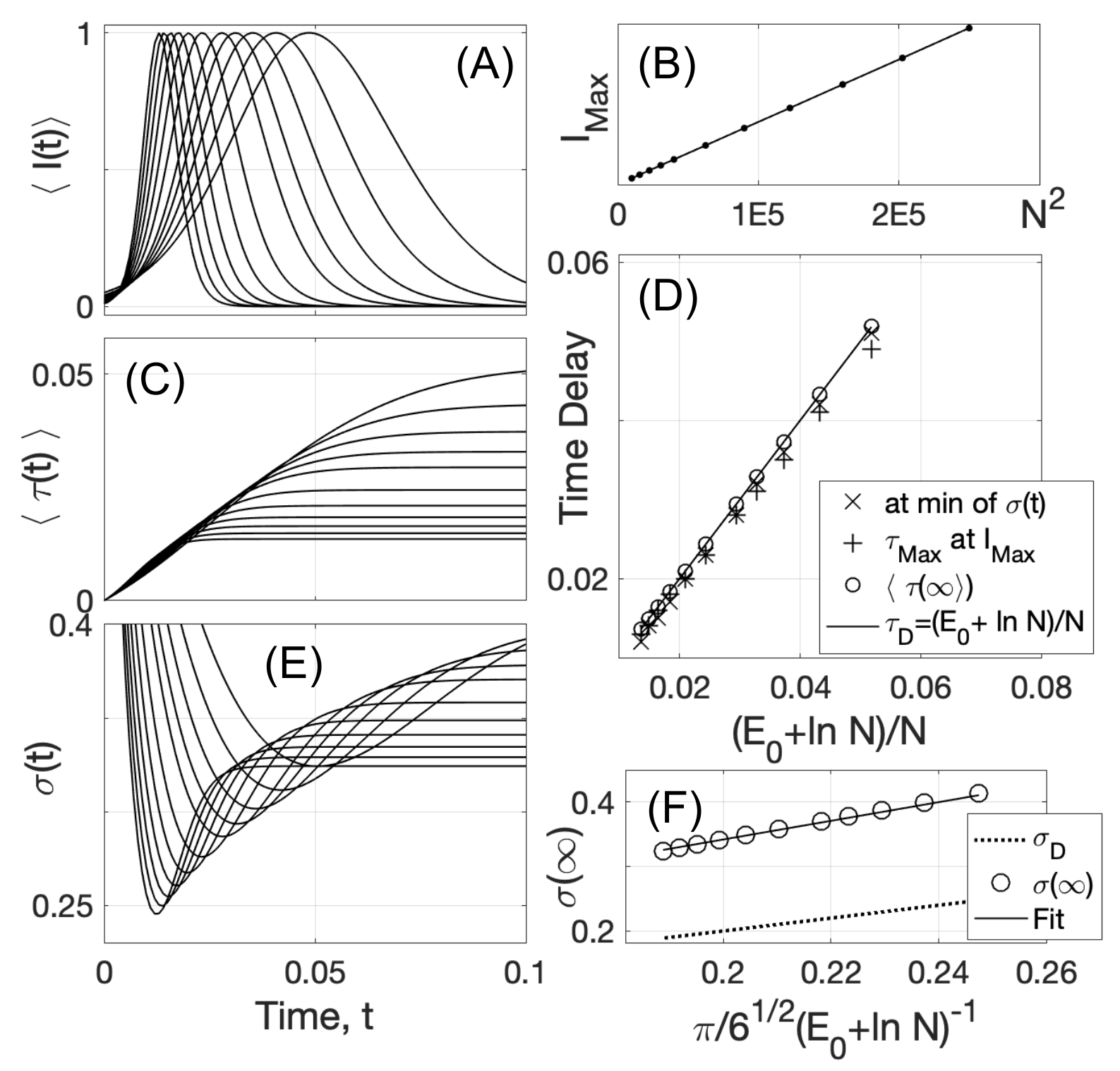}
\caption{A simple superradiance model for a $N$ two-level atomic system. (A) The normalized time-dependent intensities for the SR emissions. (B) The intensity maxima versus the number of atoms squared. (C) Time-dependent average time delays for 
different number of atoms as functions of the known analytical formula. (D) The SR time delays obtained in four different ways. (E) The normalized time-dependent fluctuations. (F) The fluctuations in the SR time delays as functions of the known analytical formula.
} 
\label{2level}
\end{center}
\end{figure}
%
%
The pulse shape i.e., time-dependent intensity is defined by~\cite{lee, agar} $\langle { I}(t)\rangle =\sum_{n=0}^{N} I(n) P_{n}(t)$ and the details are summarized in Fig.~\ref{2level}. In all plots, number of atoms is varied as $N=100$, $125$, $150$, $175$, $200$, $250$, $300$, $350$, $400$, $450$ and $500$. In Fig.~\ref{2level}(A), the pulse shapes are plotted as functions of time $t$. An important SR property is that the intensity maxima are proportional to $N^2$. That is mainly because of dominant rate term $I({N/2+1})=(N/2+1)N/2 \propto N^2$. In Fig.~\ref{2level}(B), intensity maxima as functions of $N^2$ exhibit a linear relationship. To display how the SR eventually builds up in time as number of atoms varies, the normalized pulse shapes are plotted in Fig.~\ref{2level}(A). From Fig.~\ref{2level}(A), it is shown that intensity maxima of SR are delayed farther as $N$ decreases. This is another SR behavior, which is governed by a well-known formula $\tau_D=({\rm E_0}+\ln N )/N$, where ${\rm E_0}=0.57721...$ is the Euler's constant~\cite{lee}. 

In this work, this SR time delay is defined in four different ways and all data are shown in Fig.~\ref{2level}(D). 
First, the SR time delay is defined traditionally at positions for intensity maxima $\langle I (\tau_{Max}) \rangle=I_{max}$. These time delays are plotted as functions of $({\rm E_0}+\ln N )/N$ by the lined up 'plus' marks.

Second, the SR time delay is defined by a partially averaged quantity, $\langle \tau_D \rangle_{\rm part} =\int_0^\infty t Q_{N/2+1}(t) dt$~\cite{lee}. The transition probability from a state of $N/2+1$ atoms in the upper level at $t$ to a state of $N/2$ atoms in the upper level at $t+dt$ is defined as $Q_{N/2+1}(t) dt$~\cite{lee}, with probability density $Q_{N/2+1}(t)=I({N/2+1})P_{N/2+1}(t)$. For arbitrary $n$, the density is then defined as $Q_{n}(t)=I({n})P_{n}(t)$ and all other probabilities are not included in the averaging. As found in~\cite{lee}, $\langle \tau_D \rangle _{\rm part}= \sum_{n=1}^{N/2} 1/I(n) = ({\rm E_0}+\ln N)/N$, where $N$ is large and throughout the text, we assume that it is the case. In addition, a graph $\tau_D$ versus $\tau_D$ is also plotted by a solid line as for a one-to-one comparison against the other data.

A new insight to the two-level traditional model, we adopt this idea for preparation of the precondition for a single-photon Dicke SR~\cite{ anatoly,ralf}. The transition probability from a state of a single atom in the upper level and $N-1$ atoms are in the lower level at $t$ to a state where all atoms in the lower level at $t+dt$ is $Q_{1}(t) dt$ with probability density $Q_{1}(t)=I({1})P_{1}(t)$. The average time delay for this particular condition is given by $\langle \tau^{\rm single} \rangle_{\rm part} $ $=$ $\int_0^\infty $ $t Q_{1}(t) dt$. Hence, we obtain $\langle \tau^{\rm  single} \rangle_{\rm part} $ $=$ $ \sum_{n=1}^{N} $ $1/I(n)$ $=$ $2({\rm E_0}+\ln N)/N$ $=$ $2\tau_D$. The intensity maximum is proportional to $N$ due to the dominant rate term 
$I({1})=N$ in the probability density expression. The probability for a  state to have a single excited atom is actually maximized at this time to reach $P_1(2\tau_D)$ $=$ $0.24$ ($24\%$). However, the unwanted probabilities for having two excited atoms and all de-excited atoms are $P_2(2\tau_D)$ $=$ $0.11$ ($11\%$) and $P_0(2\tau_D)$ $=$ $0.44$ ($44\%$), respectively. This means that the generation of a pure atomic Fock state relying solely on the collective spontaneous emission processes is impossible and it must require some additional procedures. For example, that may include the controlled laser excitations causing a dipole blockade effect to demonstrate with the probability of having a single excited atom up to $62\%$~\cite{fock}. The single photon Dicke SR phenomenon has no classical analog because it is unknown which particular atom is excited, thus, emitted the photon~\cite{anatoly,ralf}.

Third, let us introduce a new time-dependent expectation values of an arbitrary function $f(t)$ as 
%
%
%
\begin{equation} \label{function}
\langle  f(t)\rangle = \frac{1}{A(t)} \sum_{n=0}^{N} I({n}) \int_{0}^{t} f(t') P_{n}(t') dt'. 
\end{equation}
where all probability densities $Q_n(t)$ are contributed in overall averaging. It is normalized by an incomplete temporal pulse area $A(t)$ $=$ $\sum_{n=0}^{N}$ $I({n})$ $\int_{0}^{t}$ $P_{n}(t')$ $dt'$. 
Thus, from Eq.~(\ref{function}), time-dependent average time delay is given by $\langle \tau (t)\rangle $ $=$ $1/A(t)$ $\sum_{n=0}^{N}$ $I({n})$ $\int_{0}^{t}$ $t' P_{n}(t')dt'$. 
The temporal evolutions of average time delay for various numbers of atoms are shown in Fig.~\ref{2level}(C).  The general behavior is that the evolutions monotonically increase until they saturate. At $t=\infty$, the total temporal pulse area becomes $A(\infty)=N$. We also obtain $\sum_{n=0}^{N}$ $I({n})$ $\int_{0}^{\infty}$ $t' P_{n}(t')dt'$ $ = $   $N$ $ \langle \tau_D \rangle$. To derive this important new result, the identities $1/I(n)$ $=$ $1/I(N-n+1)$ and $\sum_{n=1}^{N}$ $\sum_{i=n}^{N}$ $1/I(i)$ $=$ $N/2$ $\sum_{i=1}^{N}$ $1/I(i)$ are used. Thus, the averaged time delay which is normalized by the total pulse area is, indeed, the SR time delay $\tau_D$
\begin{equation} \label{taularge}
\langle \tau(\infty) \rangle =\frac{1}{N} \sum_{n=0}^{N} I({n}) \int_{0}^{\infty} t' P_{n}(t') dt' = \frac{{\rm E_0}+\ln N}{N}.
\end{equation}
This is our third definition for the SR time delay. Quantities $\langle \tau(\infty) \rangle$ are numerically evaluated and the data are plotted for various numbers of atoms by the open circles in Fig.~\ref{2level}(C). The numerical result is lined up well with the analytical formula for $\tau_D$, thus, validating Eq. (\ref{taularge}). 

The fourth definition is based on the time-dependent fluctuations in time delay. An explicit analytical expression for the fluctuations for the SR time delay is well known and given by~\cite{lee,agar} as $\sigma_D$ $=$ $\pi/\sqrt{6}$ $({\rm E_0}+{\ln}N)^{-1}$. As before, time-dependent fluctuations are defined as $\sigma (t)$ $=$ $ (\langle \tau^{2}(t)\rangle $ $-$ $\langle \tau(t) \rangle^2)^{1/2}$ $/$ $\langle \tau (t) \rangle$, here $\langle \tau^{2}(t)\rangle$ $=$ $1/A(t)$ $ \sum_{n=0}^{N} $ $I({n})$ $\int_{0}^{t} $ $t'^2 P_{n}(t')dt'$ from Eq.~(\ref{function}). 
In Fig.~\ref{2level}(E), $\sigma(t)$ are plotted as functions of time $t$ for different numbers of atoms. It is important to note that the temporal evolutions of the fluctuations pass through the extreme points (minima) before they reach the stationary values at later time $t=\infty$. The minima of the time-dependent fluctuations are added to Fig.~\ref{2level}(D), see 'cross' marks. These points are in a rather good agreement with the analytical data given by $\tau_D=({\rm E_0}+\ln N )/N$. Although, in despite of tiny discrepancies in the lower region from $N=100$ to $N=200$, the agreement gets finer as $N$ increases beyond $200$. Thus, the SR time delay is defined here as the minimum position of time-dependent fluctuations $\sigma(t)$. 
Moreover, at $t=\infty$, the fluctuations are numerically shown that they are proportional to $\sigma_D$ for the fluctuations in the SR time delays $\tau_D$. The data are plotted by the open circles in Fig.~\ref{2level}(F) against $\sigma_D$ and are validated by a linear fit with a norm of residuals of $<0.01$.
%
%
%
%
%
%
%
%
%
%
%
%
%
%
%
\section{A Simple Cascade Superradiance Model}

\begin{figure}[!ht] \label{threeN}
\begin{center}
\includegraphics[width=80mm]{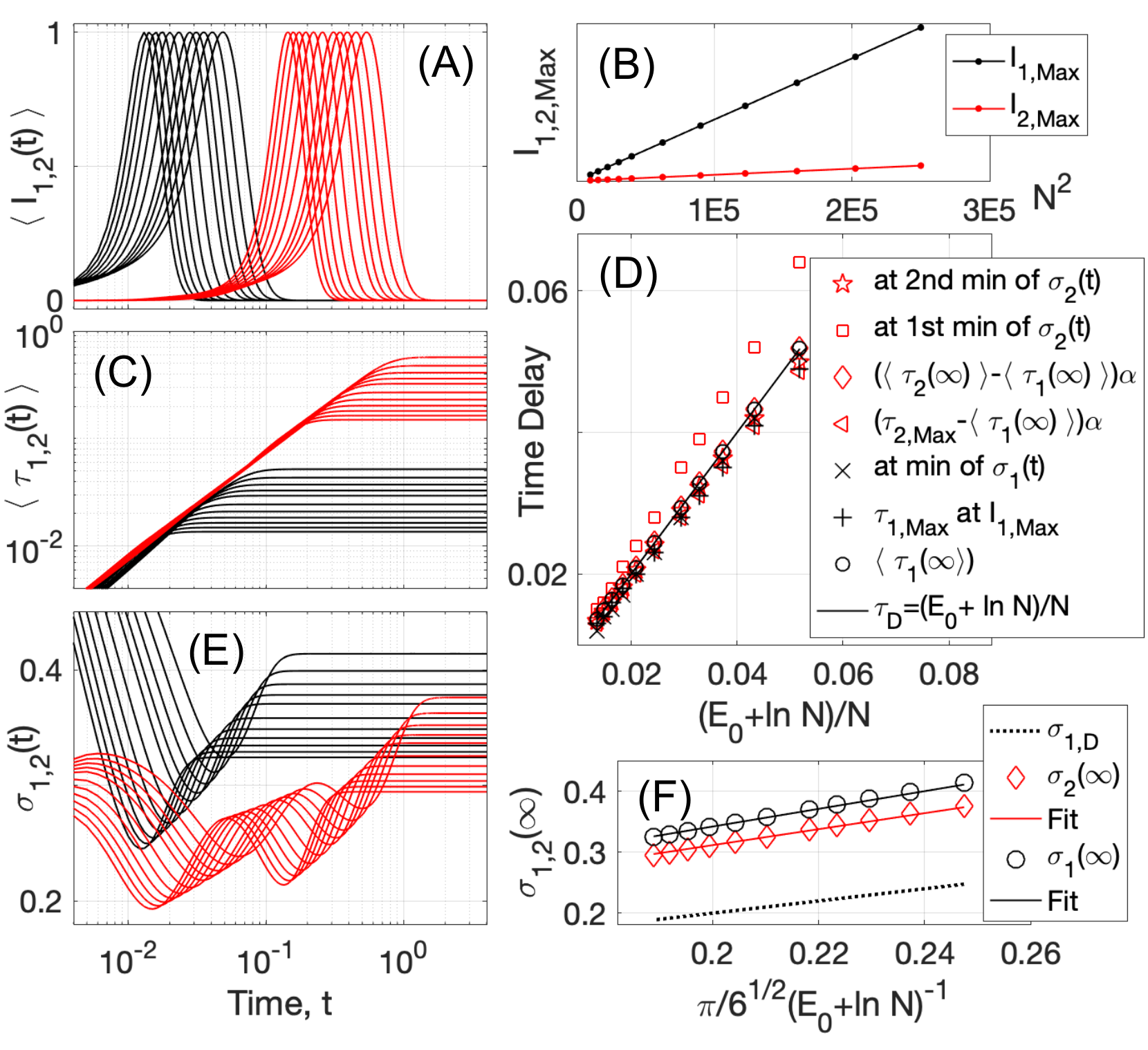}
\caption{A simple cascade superradiance model for a $N$ three-level atomic system, time axis is in log scale. (A) The normalized time-dependent intensities for pairs of SR pulses.  (B) The intensity maxima versus the number of atoms squared. (C) Time-dependent average time delays for the two modes as functions of the known analytical formula, in log-log scale. (D) The CSR time delays (scaled with $\alpha$ for the second mode) obtained in eight ways. (E) The normalized time-dependent fluctuations for the two modes. (F) The fluctuations in the CSR time delays as functions of the known analytical formula.
} 
\end{center}
\end{figure}

Next, we extend this model to a new model of two-mode cascade three-level $N$ atoms. The quantum statistical theory introduced in Ref.~\cite{agar} is adopted in this work as to reveal the main properties of collective spontaneous emissions from three-level atoms. In this case the direct two-photon transitions from/to the upper to/from lower levels are forbidden and only transitions that involve the intermediate level are allowed.  Moreover, as mentioned above, all coherence terms for individual atoms are not included in this model. Therefore, the equation of motion is reduced to~\cite{agar}
\begin{eqnarray} \label{main}
\frac{dP_{n,m}(t)}{d t} = I_1({n+1,m})P_{n+1,m}(t)-I_1({n,m})P_{n,m}(t) \nonumber\\
+ I_2({n,m+1})P_{n,m+1}(t)- I_2({n,m})P_{n,m}(t). 
\end{eqnarray}
where $P_{n,m}(t)$ is probability for a system with $n$ atoms in the upper level, $m-n$ atoms in the intermediate level, $N-n$ atoms in the lower level, i.e., for an atomic state $|n,m-n,N-n\rangle$ at time $t$. The total sum of the probability $P_{n,m}(t)$ is independent on time and must be one $\sum_{n,m=0}^{N} P_{n,m}(t) \equiv 1$. 
The cooperative decay rates, in this case, are given by $I_1({n,m})=n(m-n+1)$ for the upper transition (between the upper and intermediate levels) and $I_2({n,m})=\alpha (m-n)(N-m+1)$ for the lower transition (between intermediate and lower levels)~\cite{agar}. $\Gamma_{1}$ and $\Gamma_{2}$ are the spontaneous decay rates for the upper and lower transitions of a single atom, respectively. Similarly, $\mu_{1,2}$ are geometrical factors~\cite{feld2} and a dimensionless time $t$ is introduced as $t \rightarrow \mu_1\Gamma_1 t$. A new scaling quantity is given by $\alpha={\mu_2\Gamma_{2}}/{\mu_1\Gamma_{1}}$, here we consider $\alpha < 1$. 
For example for rubidium atoms the ratio is $\alpha=0.01$ for the two-photon transition between $5S$ ground state and $5D$ excited state via $6P$ intermediate state. The upper transition (lower) corresponds to $5 {\rm \mu m}$ ($420$ nm)~\cite{ari1,ariOL,ariFluct,ariBack}. Similarly, for cesium atoms, the two-photon transition is between $6S$  ground state and $8S$ excited state via either $7P$~\cite{ariCs} or $6P$~\cite{brownell} intermediate state. For sodium atoms, that is between $3S$ ground state and $4S$ excited state~\cite{ariNa}. 
The characteristics for the first SR pulse are identical to those obtained in the above for the two-level atoms. However, the characteristics for the secondary SR pulse is slightly modified as explained later. In Fig.~\ref{threeN}, the CSR quantities are plotted as $N$ varies (with $\alpha=0.1$) in the similar format as shown in Fig.~\ref{2level}. The time-dependent intensities in the two transitions can be characterized by~\cite{agar}
$ \langle {I}_{1,2}(t)\rangle =\sum_{n,m=0}^{N} I_{1,2}({n,m}) P_{n,m}(t)$.
According to Eq.(\ref{main}), the time evolutions are bunched in two distinct regions with scaled times: $t$ and  $\alpha t$.  The normalized pulse shapes are shown in Fig.~\ref{threeN}(A) by solid curves, where the first bunch (black) and the trailing bunch (red) correspond to the first and secondary SR emissions. Fig.~\ref{threeN}(B) shows linear relationships of the intensity maxima of the two modes as functions of $N^2$. 
%
%
Because of experiencing two distinctly scaled times, the two pulse temporal areas are needed to normalize to obtain time-dependent average time delays as
\begin{equation} \label{tau12}
\langle \tau_{1,2}(t)\rangle =\frac{1}{A_{1,2}(t)}{\sum_{n,m=0}^{N}I_{1,2}(n,m)\int_{0}^t t' P_{n,m}(t') dt'}
\end{equation}
here the partial pulse areas are $A_{1,2}(t)$ $=$ $\sum_{n=0}^{N}$ $I_{1,2}({n})$ $\int_{0}^{t}$ $P_{n}(t)dt$. The average time delays are plotted in Fig.~\ref{threeN}(C), in a log-log scale. As before, the SR time delays are the expectation values determined at $t=\infty$, $\langle \tau_{1}(\infty) \rangle$ $=$ $\tau_{1D}$. This statement is supported by the fact that the first SR emissions are not affected by the initiation of the secondary SR emissions. The first mode is mainly governed by Eq.~{\ref{main2level}} for the two-level model. However, for the second mode, its time delay is $\langle \tau_{2}(\infty) \rangle$ $\simeq \tau_{1D}$ $+$ $\tau_{2D}$. It is also numerically shown that $\tau_{2D}$ $ \simeq$ $ \tau_{1D}/ \alpha$, thus we obtain  
\begin{eqnarray} \label{tau12D}
\langle \tau_{1}(\infty) \rangle & \simeq & \tau_{1D}\nonumber\\
\langle \tau_{2}(\infty)\rangle -\langle \tau_{1}(\infty) \rangle & \simeq & \tau_{2D} \simeq \frac{\mu_1\Gamma_1}{\mu_2\Gamma_2}\tau_{1D}
 \end{eqnarray}
These results are plotted as functions of $({\rm E_0}+\ln N )/N$ in Fig.~\ref{threeN}(D). The linear relationships are displayed for both $\langle \tau_1(\infty) \rangle$ and $\langle \tau_2(\infty) \rangle$. 
The same marks as in Fig.~\ref{2level} are used in the graph for first SR pulse data in black. The data for $\langle \tau_{2}(\infty) \rangle$ are marked by open diamonds in red.  Analogously, the time-dependent normalized fluctuations are defined as $\sigma_{1,2}(t)$ $=$ $(\langle\tau_{1,2}^{2}(t)\rangle$ $-$ $\langle \tau_{1,2} (t)\rangle^2)^{1/2}$ $/\langle \tau_{1,2} (t)\rangle$, here $\langle \tau_{1,2}^{2}(t)\rangle$ $=$ $1/A_{1,2}(t)$ $ \sum_{n,m=0}^{N} $ $I_{1,2}({n,m})$ $\int_{0}^{t} $ $t'^2 P_{n,m}(t')dt'$. In this case, the fluctuations for the second mode exhibit two minima: one as expected at $\tau_{2D}$, but another near $\tau_{1D}$. This is one of the advantages of using the time dependent quantities as it reveals an occurrence of the triggered (or yoked) emissions for the two modes. However, in contrary to the known yoked-emissions~\cite{brownell,ari1,ariNa,ariCs}, no coherence of any individual atom is present here. These minima are plotted in Fig.~\ref{threeN}(D) by open squares in red, which are lined up with the differently defined time delays.
\section{Discussions}
\begin{figure}[!ht] 
\begin{center}
\includegraphics[width=80mm]{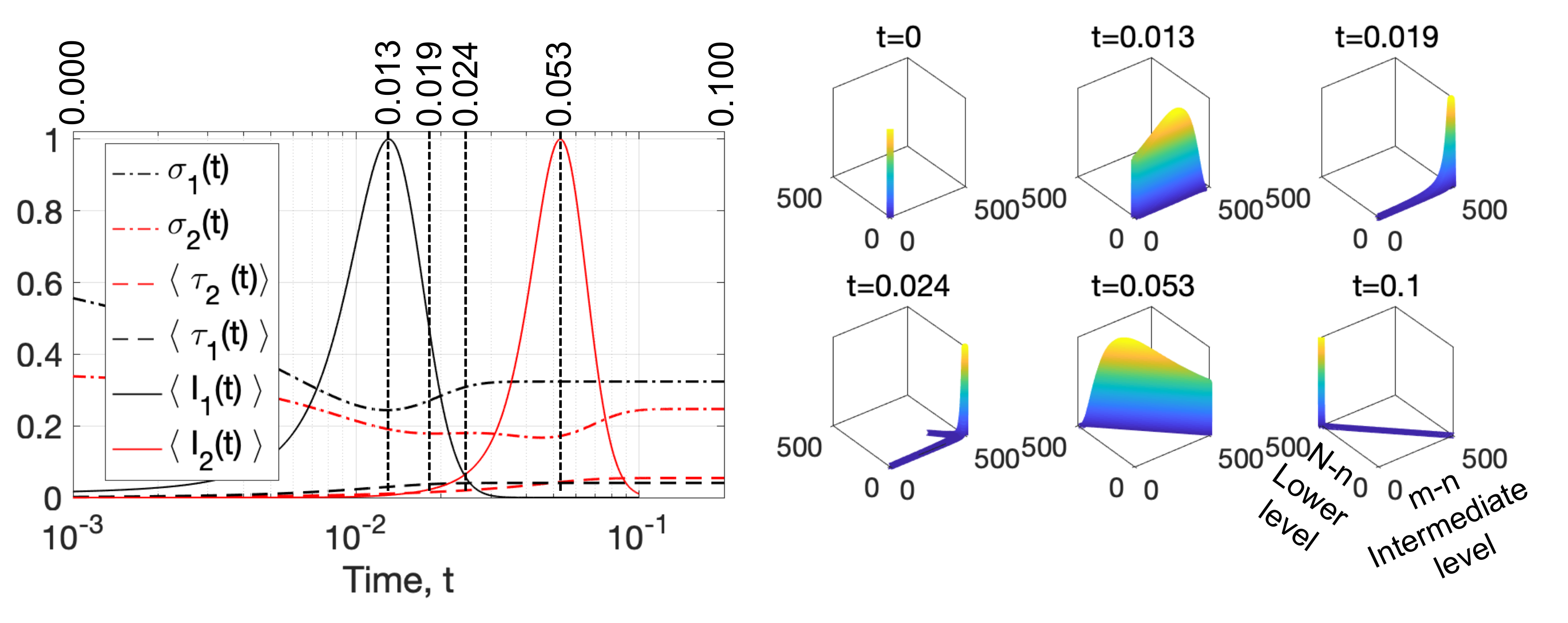}
\caption{Left: The normalized CSR pulse shapes as functions of time $t$ (time axis in a log scale). Right: The temporal evolution of the atomic state probabilities shown for several selected times.} 
\label{sample}
\end{center}
\end{figure}

In this section, a simple CSR model is explained in detail for the case of $N=500$ atoms and $\alpha = 1/3$. In Fig.~\ref{sample}, left panel, the normalized time-dependent intensities (solid curves), time delays (dot-dashed curves)  and fluctuations (dashed curves) are plotted by  black (first mode) and red (second mode). Dotted lines correspond to the selected times, which divide a whole interval into several regions. At $t=0$, we assume that the pumping by ultrashort pulses prepares all $N$ atoms only in the upper level from the off-resonant ground state but not the intermediate or lower level. The pumping is ultrafast before any atomic decay processes in the upper and lower transitions take place. At $t=$ $\tau_{1D}$ $=0.013$, the first SR emissions in the upper transition occurs. At this time, intensity reaches its maximum. Coincidently at $t=\tau_{1D}$, the fluctuations and time delay reach the minimum and stationary value $\langle \tau_1 (\infty)\rangle$, respectively. In the next region between $t=0.013$ and $0.024$, it is important to note that the first SR emissions in the upper transition trigger an initiation of the emissions in the lower transition. At $t=0.019$, the fluctuations in time delays for the secondary mode reaches its first minimum. Then, at $t=0.024$, a macroscopic mixture of the two mode states are generated. These emissions in both transitions occur at the same time, therefore, are strongly correlated. It is remarkable that these two-mode correlated emissions are originated from the same group of the atoms rather than some superposed sub-ensembles. 
%
Later, at $t=0.053$, the intensity for the second mode  reaches its maximum, i.e., the secondary SR in the lower transition occurs. This time is consistent with Eq.~(\ref{tau12D}), where $\langle \tau_2 (t) \rangle$ $ \simeq$ $(1+1/\alpha)$  $\tau_{1D}$ $=$  $0.052$ and $\tau_{2D}$ $=$ $\tau_{1D}$ $/$ $\alpha$ $=$ $0.013$, for given $\alpha=1/3$. Let us estimate the CSR results for the rare-earth doped material used in the recent CSF experiment~\cite{chiossi}. As pointed out in~\cite{chiossi}, the non-radiative decay processes may be involved in the CSF, that is not the case for CSR. We assume free of non-radiative decay process. The spontaneous decay rates in both the upper (2718 nm) and lower (1534 nm) transitions are chosen to be the same, 14.5 s$^{-1}$~\cite{chiossi}, and $\mu_{1,2}$ are defined as in~\cite{feld2}. The CSR parameter is estimated to be $\alpha\simeq 1/3$ for this experimental arrangement. Thus, the secondary SR time delay is about three times longer than the first SR time delay. We estimate the first SR time delay is about $0.2$ $\mu$s by looking at the interval between the peaks is about $0.7$ $\mu$s from the data shown  in Fig. 3 in~\cite{chiossi}, assuming free of non-radiative decay process.
Finally, at $t=0.1$ all atoms are in the lower state and none is left in the upper or intermediate level and all behaviors become stationary (or zero).
In the left panel of Fig.~\ref{sample}, the temporal evolutions of the probability are displayed. The corresponding probabilities $P_{n,m}(t)$ at those selected times are depicted as functions of $N-n$ (number of atoms in the lower level) and $m-n$ (number of atoms in the intermediate level) rather than $n$ and $m$. In addition, $P_{n,m}(t)$ can also be displayed as functions of $n$ (number of atoms in the upper level) and either $N-n$ or $m-n$, not shown here. 
%
%
\section{Conclusion}
In conclusion, a simple cascade superradiance model is developed in this work. Temporal evolutions of average time delays and the fluctuations are introduced. These formulations allow to obtain the superradiance time delays and fluctuations in several different ways. The main characteristics of the correlated two-mode emissions are obtained. In the future, these correlated emissions may be used, for example, in quantum noise quenching. Previously, quantum fluctuations were measured by simultaneously generating a pair of superfluorescence pulses from the two separate sub-ensembles of collective atoms~\cite{vrehen, ariFluct}. However, the cascade multi-level excitations may provide much stronger correlated cooperative emissions originated from a single ensemble of collective atoms.   
{\bf Acknowledgement}
We would like to thank Prof. Girish Agarwal and Prof. Marlan Scully for their stimulating discussions.
\end{document}